\documentclass[aps,prl,preprint,superscriptaddress]{revtex4-2}

\usepackage{graphicx}
\usepackage{natbib}

\usepackage{dcolumn}
\usepackage{amsmath}
\usepackage{braket}
\usepackage{float}
\usepackage{bm}
\usepackage[english]{babel}

\makeatletter
\def\btt#1{\texttt{\@backslashchar#1}}
\DeclareRobustCommand\bblash{\btt{\@backslashchar}}
\makeatother

\usepackage{color}

\begin{document}

\preprint{}

%%%%%%%%%%%%%%%%%%%%%%%%%%%%%%%%%%%%%%%%%%%%%%%%%%%%%%%%%%%%%%%%%%%%%%%%%%%%%%%

\title[Short Title]{Microwave hinge states\\
in an all-dielectric woodpile photonic crystal insulator}

\author{Shun Takahashi}
\affiliation{
Kyoto Institute of Technology, Sakyo, Kyoto 606-8585, Japan
}
\affiliation{
Graduate School of Engineering, Tohoku University, Aoba, Sendai, Miyagi 980-8579, Japan
}

\author{Yuya Ashida}
\affiliation{
Kyoto Institute of Technology, Sakyo, Kyoto 606-8585, Japan
}

\author{Huyen Thanh Phan}
\affiliation{
Department of Nanotechnology for Sustainable Energy, School of Science and Technology, Kwansei Gakuin University, Gakuen Uegahara, Sanda 669-1330, Japan
}
\affiliation{
Center for Materials Innovation and Technology, VinUniversity, Hanoi 100000, Vietnam
}

\author{Shuichiro Kamei}
\affiliation{
Kyoto Institute of Technology, Sakyo, Kyoto 606-8585, Japan
}

\author{Kenichi Yamashita}
\affiliation{
Kyoto Institute of Technology, Sakyo, Kyoto 606-8585, Japan
}

\author{Tetsuya Ueda}
\affiliation{
Kyoto Institute of Technology, Sakyo, Kyoto 606-8585, Japan
}

\author{Katsunori Wakabayashi}
\affiliation{
Department of Nanotechnology for Sustainable Energy, School of Science and Technology, Kwansei Gakuin University, Gakuen Uegahara, Sanda 669-1330, Japan
}
\affiliation{
Research Center for Materials Nanoarchitectonics, National Institute for Materials Science, Namiki, Tsukuba 305-0044, Japan
}

\author{Satoshi Iwamoto}
\affiliation{
Research Center for Advanced Science and Technology, The University of Tokyo, Meguro, Tokyo 153-8505, Japan
}
\affiliation{
Institute of Industrial Science, The University of Tokyo, Meguro, Tokyo 153-8505, Japan
}

%%%%%%%%%%%%%%%%%%%%%%%%%%%%%%%%%%%%%%%%%%%%%%%%%%%%%%%%%%%%%%%%%%%%%%%%%%%%%%%

\date{\today}

%%%%%%%%%%%%%%%%%%%%%%%%%%%%%%%%%%%%%%%%%%%%%%%%%%%%%%%%%%%%%%%%%%%%%%%%%%%%%%%

\begin{abstract}
Higher-order topological states are studied in a woodpile photonic crystal (PhC) having a large complete photonic bandgap. 
According to theoretical predictions, we fabricate a three-dimensional all-dielectric PhC with a millimeter-scale lattice period. 
The structure consists of two distinct lattice types that coincide through a structural displacement, where their interface intersection gives rise to hinge states. 
Microwave transmission measurements reveal transmission peaks and spatial localization at the intersection, demonstrating the hinge states. 
Furthermore, we confirm the absence of hinge states at a specific intersection. 
As theoretically predicted, hinge states emerge at intersections where the localized boundary polarizations are topologically different, whereas they are absent at intersections where the boundary polarizations are equivalent. 
\end{abstract}

%%%%%%%%%%%%%%%%%%%%%%%%%%%%%%%%%%%%%%%%%%%%%%%%%%%%%%%%%%%%%%%%%%%%%%%%%%%%%%%

\maketitle

%%%%%%%%%%%%%%%%%%%%%%%%%%%%%%%%%%%%%%%%%%%%%%%%%%%%%%%%%%%%%%%%%%%%%%%%%%%%%%%
\section{Introduction}
%%%%%%%%%%%%%%%%%%%%%%%%%%%%%%%%%%%%%%%%%%%%%%%%%%%%%%%%%%%%%%%%%%%%%%%%%%%%%%%
Hinge states, which are second-order topological states in three-dimensional (3D) systems \cite{BBH1,Langbehn,Song,Schindler}, can serve as one-dimensional (1D) waveguides in photonics \cite{Kim,Xie,Lin1}. 
Owing to their topological protection, these waveguides can potentially suppress light scattering induced by defects or fabrication imperfections in 3D structures \cite{WangPRB}, even for optical modes above the light line \cite{Wang_MW_hinge}. 
Photonic hinge states have been experimentally demonstrated mainly in the microwave (MW) regime, including helical hinge states in a topological insulator \cite{Wang_MW_hinge} and semimetals \cite{microwave_hinge_PRB,microwave_hinge_OL,Pan}, chiral hinge states in insulators \cite{GGLiu,Lai_MW_hinge,MengXiao}, and antichiral hinge states in semimetals \cite{Peng_antichiral_hinge,Wang_antichiral_hinge}. 
However, these 3D structures involve complex geometries and contain metal or magnetic materials, which lead to substantial losses and thereby limit their practical applications. 

%%%%%%%%%%%%%%%%%%%%%%%%%%%%%%%%%%%%%%%%%%%%%%%%%%%%%%%%%%%%%%%%%%%%%%%%%%%%%%%
One of the simplest approaches to realize hinge states in real-space 3D photonic crystal (PhC) insulators composed of a single dielectric material is to introduce a structural dislocation in a part of the 3D structures. 
Such a structural shift modifies the unit-cell configuration and can provide a distinct topological invariant, such as a Zak phase \cite{Zak}, a winding number of a Wilson loop \cite{Wilson}, a bulk polarization \cite{BBH2}, or Wannier-center position \cite{Ezawa}, relative to that of the original unit cell. 
We have previously demonstrated MW hinge states in a 3D all-dielectric PhC insulator based on a simple cubic lattice \cite{TakahashiPRB}. 
The structure consists of two distinct lattice types that coincide through a structural displacement, where their interface intersection gives rise to hinge states. 
Unlike hinge states at air interfaces \cite{Wang_MW_hinge}, these hinge states are embedded in the PhC insulator with a complete photonic bandgap (CPBG) \cite{Yab,Joanno}. 

%%%%%%%%%%%%%%%%%%%%%%%%%%
However, in the simple-cubic-lattice PhC, the relatively small CPBG causes the hinge-mode dispersion to overlap spectrally with that of first-order topological boundary modes in the projected band structure. 
All-dielectric PhCs based on diamond lattices \cite{Ho1} are known to have the largest CPBGs among dielectric lattices across all space groups \cite{Men,Watanabe}. 
Such a wide bandgap can spectrally isolate hinge modes from other topological boundary modes. 
%Such a large bandgap enables defect or boundary modes to be spectrally isolated, which is advantageous for optical characterization and device implementation. 
As a simplified realization of the diamond lattice, the woodpile structure has been devised by dividing the lattice into four layers and representing each layer as an array of dielectric rods \cite{Ho2,Maldovan}. 
We have theoretically investigated hinge states in a woodpile structure and identified spectral regions in which a part of the hinge-mode dispersion is isolated from other boundary modes \cite{Phan}. 
Nevertheless, their experimental verification has not yet been achieved. 

%%%%%%%%%%%%%%%%%%%%%%%%%%%%%%%%%%%%%%%%%%%%%%%%%%%%%%%%%%%%%%%%%%%%%%%%%%%%%%%
In this study, we experimentally demonstrate hinge states in the MW regime, based on the theoretical prediction. 
We fabricate a millimeter-scale, all-dielectric 3D PhC insulator based on a woodpile structure. 
The 3D structure consists of two distinct lattice types related by a structural displacement. 
The interface between these domains hosts boundary modes, and their intersection gives rise to hinge-localized states. 
We perform MW transmission measurements on the 3D structure and observe transmission peaks and spatial localization of the hinge states, in good agreement with numerical simulations. 
In addition, we experimentally show the absence of hinge states at specific intersections. 
This is consistent with the theoretical prediction \cite{Phan} that hinge states emerge at intersections where the localized boundary polarizations are topologically different, whereas no hinge states are expected at intersections where the boundary polarizations are equivalent. 
The observed hinge states, which are strongly confined by the CPBG, offer potential applications for low-loss, metal-free MW transmission lines. 
Moreover, by scaling down the lattice period of the studied PhCs to the sub-$\mu$m range, hinge states operating in the near-infrared regime could be realized using 3D nanofabrication techniques, such as a micro-manipulation method \cite{Aoki,Aniwat,TakahashiEL,TakahashiAPEX}. 

%%%%%%%%%%%%%%%%%%%%%%%%%%

\begin{figure}[t]
\centering\includegraphics[width=0.9\linewidth]{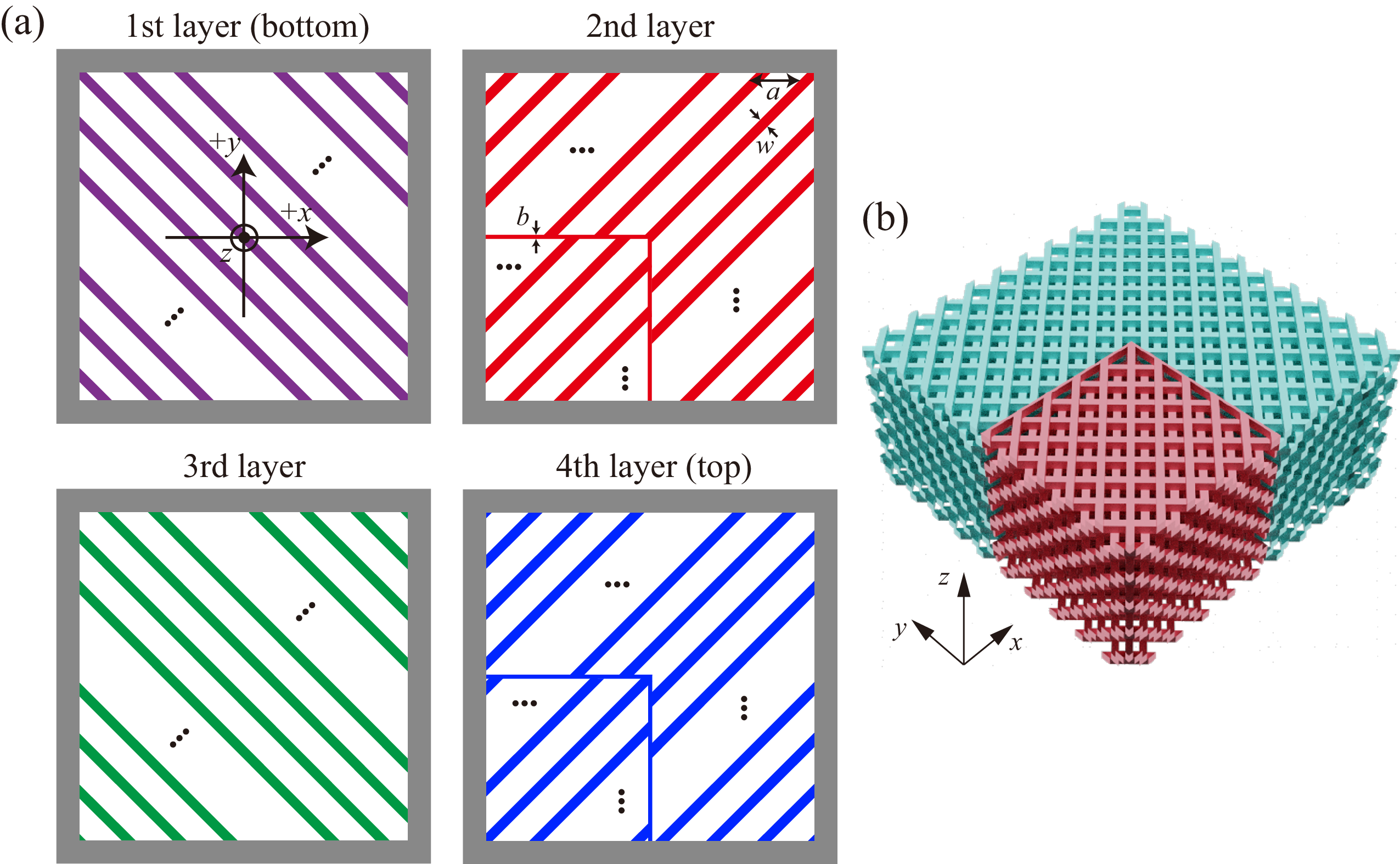}
\caption{
(a) Schematic diagrams of the four layers with frames. 
In the second and fourth layers, one quarter of the pattern is laterally shifted by $a$/2, which is surrounded by beams and frames to be self-suspended. 
The coordinates and the structural parameters are shown in the first and second layer, respectively. 
(b) Schematic diagram of the stacked 3D structure using the four layers shown in (a). 
The frames are removed for clarity. 
The magenta and cyan regions correspond to domains with distinct topological invariants. 
}
\end{figure}

%%%%%%%%%%%%%%%%%%%%%%%%%%%%%%%%%%%%%%%%%%%%%%%%%%%%%%%%%%%%%%%%%%%%%%%%%%%%%%%
\section{Design of the woodpile photonic crystal}
%%%%%%%%%%%%%%%%%%%%%%%%%%%%%%%%%%%%%%%%%%%%%%%%%%%%%%%%%%%%%%%%%%%%%%%%%%%%%%%
Figure 1(a) shows the four layers of the studied layer-by-layer structure based on a woodpile structure. 
Each layer has a line-and-space pattern with an in-plane period of $a$ = 12.8 mm and a thickness of $a$/4. 
The colored lines/rods with a width of $w$ = 0.2$a$ are dielectrics with a refractive index of 3.5. 
Adjacent layers have orthogonally oriented rod arrays, whereas next-nearest layers contain parallel rods shifted by $a$/2 along either the $x$- or $y$-direction. 
Numerical calculation using a finite element method shows that this simple woodpile structure forms a CPBG from 11.5 GHz to 13.7 GHz. 

%%%%%%%%%%%%%%%%%%%%%%%%%%
According to the previous theoretical report \cite{Phan}, a structural shift of the woodpile structure by ($\Delta x$, $\Delta y$, $\Delta z$) = ($a$/4, $a$/4, $a$/2) provides different topological invariants. 
This 3D shift corresponds to the $a$/2 shifts in the $x$- or $y$-direction in the second and fourth layers. 
In this study, when the origin of the $x$-$y$ coordinate is defined at the center of each layer, these shifts are introduced in the (-$x$, -$y$) quarter part of the second and fourth layers, as shown in Fig. 1(a). 
Hinge states can originate from a mismatch of higher-order bulk topological invariants across structural domains. 
This specific relative shift in the woodpile lattice modifies the bulk polarization, whose topological difference is characterized by the winding numbers of the Wilson loop spectrum \cite{Phan}, leading to localized boundary polarization at the intersection of two topologically distinct surfaces. 
This bulk-hinge correspondence provides the fundamental mechanism for the emergence of hinge modes.

%%%%%%%%%%%%%%%%%%%%%%%%%%
To ensure mechanical stability in the following fabrication processes, a frame is incorporated around the pattern in each layer, and additional supporting beams are introduced along the boundaries in the second and fourth layers. 
Both the frame and the beams are the same dielectric as the rods, and the width of the beams is $b$ = 0.08$a$. 
The number of periods depends on the specific numerical simulation or the experimental measurement due to fabrication limitation. 
Figure 1(b) shows a schematic diagram of the stacked 3D structure with the frames removed for clarity. 
The different colors represent structural domains characterized by different topological invariants. 

%%%%%%%%%%%%%%%%%%%%%%%%%%

\begin{figure}[t]
\centering\includegraphics[width=0.9\linewidth]{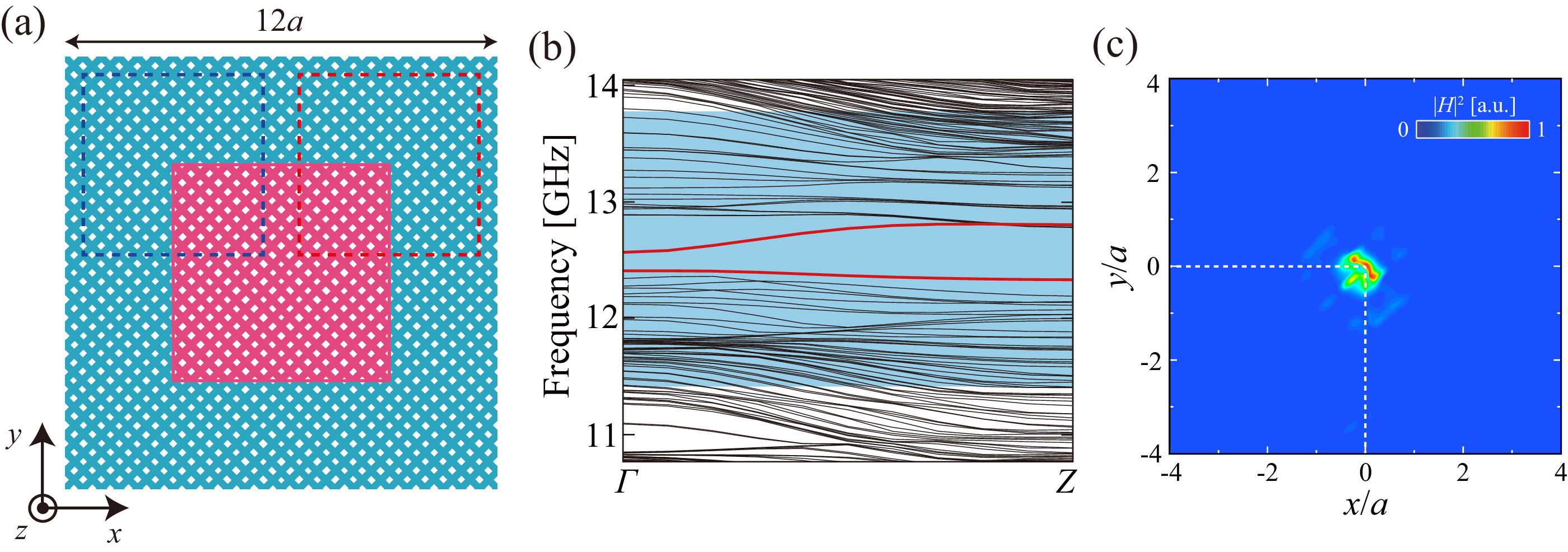}
\caption{
(a) Schematic diagram of a supercell. 
Topological invariants are different in the magenta and cyan colored domains. 
The beams in Fig. 1 are included at the interfaces of the two domains. 
Hinge states appear in the red-dotted square, whereas they are absent in the blue-dotted square. 
(b) Numerically obtained photonic band structure for the supercell in (a). 
(c) Numerically obtained spatial distribution of $|H^2|$ at 12.4 GHz. 
The $z$-position of the plotted slice corresponds to the center of the fourth layer in Fig. 1(a). 
}
\end{figure}

%%%%%%%%%%%%%%%%%%%%%%%%%%%%%%%%%%%%%%%%%%%%%%%%%%%%%%%%%%%%%%%%%%%%%%%%%%%%%%%

\section{Numerical investigation of hinge states}

%%%%%%%%%%%%%%%%%%%%%%%%%%%%%%%%%%%%%%%%%%%%%%%%%%%%%%%%%%%%%%%%%%%%%%%%%%%%%%%

First, we numerically investigate hinge states in a photonic band structure of the 3D PhC composed of the two lattice types that coincide through the shift. 
We prepare a supercell with an in-plane size of a 12$a$ square, in which a shifted structure with an in-plane size of a 6$a$ square is embedded, as shown in Fig. 2(a). 
The beams in the second and fourth layers are included at the interfaces of the two domains, which were not considered in the previous report \cite{Phan}. 
In the $z$-direction, four layers are stacked. 
Periodic boundary conditions are imposed in all three directions. 
For this supercell, we calculate the photonic band structure where all bands are projected in the $\Gamma$-$Z$ direction, using a finite element method. 
Figure 2(b) shows the projected band structure around the bandgap indicated as a blue-shaded region. 
Bands outside the bandgap correspond to bulk modes. 
The structural shift of $(a/4, a/4, a/2)$ modifies the bulk polarization configuration while preserving the CPBG. 
As demonstrated in \cite{Phan}, the Wilson loop spectrum exhibits distinct winding behaviors in the two domains. 
The associated winding numbers, corresponding to partial Chern numbers, provide the topological invariants distinguishing the two domains, which gives rise to higher-order topological phases via bulk-boundary correspondence. 
The interface between these domains therefore hosts boundary modes, and their intersection gives rise to hinge-localized states. 
In Fig. 2(b), the black bands in the bandgap correspond to boundary modes, whereas the red bands are attributed to hinge modes. 
In particular, parts of the hinge modes are spectrally isolated and therefore do not couple to the radiative continuum. 

%%%%%%%%%%%%%%%%%%%%%%%%%%

\begin{figure}[t]
\centering\includegraphics[width=0.9\linewidth]{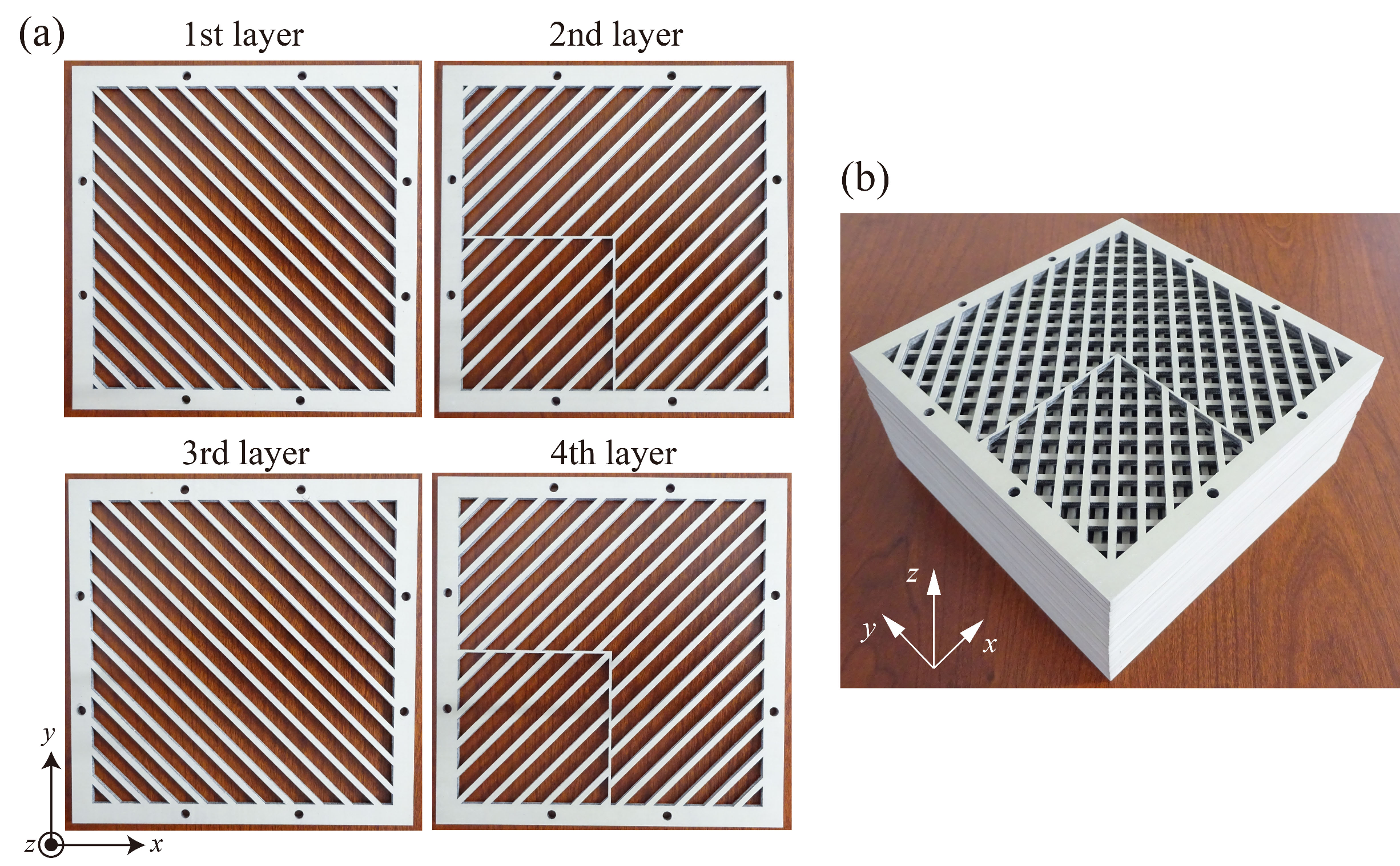}
\caption{
(a) Photographs of the fabricated four layers with the frames and beams. 
(b) Photograph of the stacked structure. The number of layers is totally 20, corresponding to five periods in the $z$-direction. 
}
\end{figure}

%%%%%%%%%%%%%%%%%%%%%%%%%%
In this study, the supporting beams are incorporated in the second and fourth layers to form self-suspended patterns, effectively introducing structural perturbations for the hinge states. 
Although this structural modulation modifies the frequencies and dispersion curves of the hinge modes, it does not annihilate them. 
In this regard, the hinge modes show some degree of robustness against such perturbations. 

%%%%%%%%%%%%%%%%%%%%%%%%%%
We further employ a finite-difference time-domain (FDTD) method to calculate the spatial distribution of $|H|^2$ at the intersection in the (+$x$, +$y$) region, indicated by the red-dotted square in Fig. 2(a). 
We prepare a 3D structure with an in-plane area of 8$a$ square, including the intersection at its center. 
The height of the structure in the $z$-direction is $a$. 
Perfectly matched layers are applied at the in-plane boundaries, while periodic boundary conditions are imposed in the $z$-direction. 
At 12.4 GHz, a spatial map of magnetic field intensity $|H|^2$ is plotted in Fig. 2(c), corresponding to a slice through the center of the fourth layer shown in Fig. 1(a). 
The intensity is clearly localized at the intersection, as indicated by the white dotted lines, confirming the confinement of the hinge state. 

%%%%%%%%%%%%%%%%%%%%%%%%%%
Note that the hinge modes appear at only two of the four intersections located in the ($\pm x$, $\pm y$) regions shown in Fig. 2(a). 
Namely, hinge states are observed at the intersection in the red-dotted square, whereas they are absent in the blue-dotted square. 
This selection rule originates from the difference in the Wilson loop on the $k_y$-$k_z$ and $k_x$-$k_z$ planes between the two structural domains \cite{Phan}. 
Owing to this difference in higher-order bulk topology, hinge modes are prohibited at the ($\pm x$, $\mp y$) intersections.
This absence of hinge modes is also demonstrated in the following MW measurements. 

%%%%%%%%%%%%%%%%%%%%%%%%%%%%%%%%%%%%%%%%%%%%%%%%%%%%%%%%%%%%%%%%%%%%%%%%%%%%%%%
\section{Microwave measurements on hinge states}
%%%%%%%%%%%%%%%%%%%%%%%%%%%%%%%%%%%%%%%%%%%%%%%%%%%%%%%%%%%%%%%%%%%%%%%%%%%%%%%

\begin{figure}[t]
\centering\includegraphics[width=0.7\linewidth]{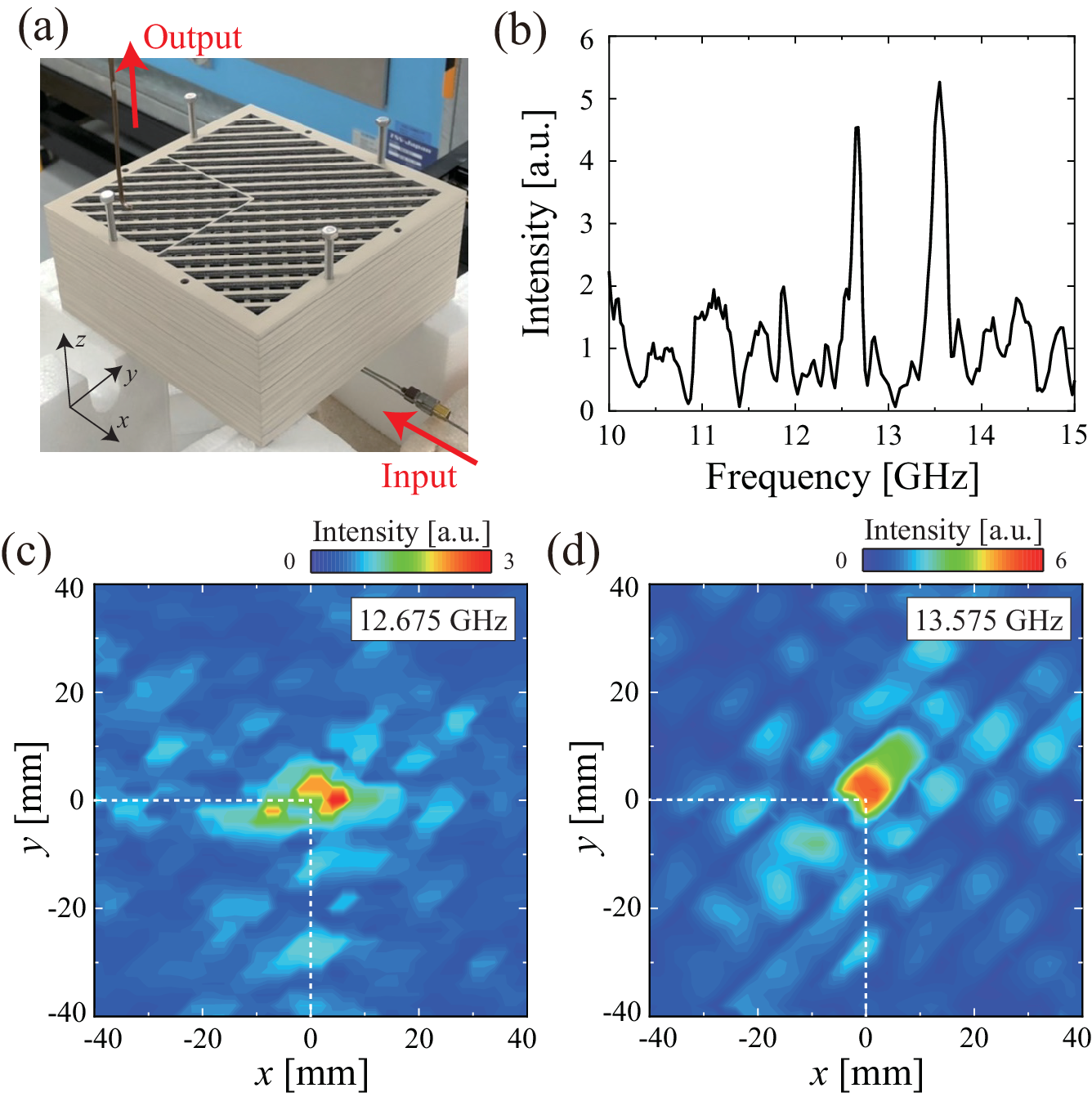}
\caption{
(a) Photograph of the measurement setup. 
One loop antenna is placed beneath the surface, while a second loop antenna, detecting the $H_z$ component of the transmitted MWs, is mechanically scanned on the top surface. 
(b) Measured transmission spectrum at the center of the top surface. 
Two intensity peaks at 12.675 GHz and 13.575 GHz correspond to the hinge states. 
(c), (d) Color maps of the intensity distribution at 12.675 GHz and 13.575 GHz, respectively. 
Intensity localization is clearly observed at the center of the top surface. 
}
\end{figure}

%%%%%%%%%%%%%%%%%%%%%%%%%%

\begin{figure}[t]
\centering\includegraphics[width=0.9\linewidth]{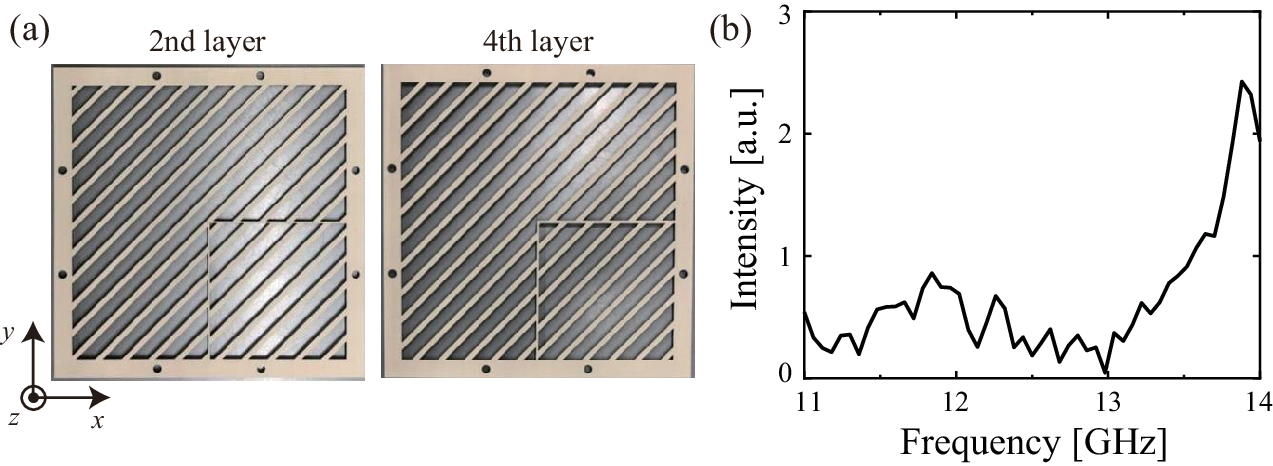}
\caption{
(a) Photographs of the two layers constructing the other intersection. 
(b) MW transmission spectrum at the center of the top surface. 
}
\end{figure}

%%%%%%%%%%%%%%%%%%%%%%%%%%
Figure 3(a) shows the four layers fabricated by drilling polytetrafluoroethylene boards with a thickness of 3.2 mm, a refractive index of 3.5 ($\pm$0.1), and a dissipation factor of $\sim$0.002 at 10 GHz. 
The lattice constant of the woodpile structure is $a$ = 12.8 mm, the rod width is $w$ = 2.6 mm, and the beam width is $b$ = 1.0 mm with minor fabrication errors of $\pm$0.2 mm in-plane and $\pm$0.5 mm out-of-plane. 
The numbers of in-plane and out-of-plane periods are 10 and 5, respectively. 
The frames have holes with a diameter of 3.3 mm to align the stacked layers. 
The 3D stacked structure is shown in Fig. 3(b). 

%%%%%%%%%%%%%%%%%%%%%%%%%%
For the fabricated 3D structure, we perform MW transmission measurements in the $z$-direction using a network analyzer (N5224A, Agilent Technologies). 
As an incident magnetic dipole, a loop antenna with a diameter of 3 mm is placed at the center of the bottom surface. 
The magnetic dipole is oriented to the $z$-direction. 
To measure the spatial distribution of transmitted MWs on the top surface, a second loop antenna is mechanically scanned in the $x$-$y$ plane with a spatial resolution of 2 mm, as shown in Fig. 4(a). 
The second loop antenna is parallel to the incident loop antenna to detect the $H_z$ component. 
From the network analyzer, we obtain the scattering parameter $S_{21}$ between the incident (port 1) and detection (port 2) antennas in the frequency range of 10-15 GHz with a resolution of 0.025 GHz. 
During the measurements, absorber sheets with 23 dB absorption at 10 GHz are attached to the side surfaces of the PhC. 

%%%%%%%%%%%%%%%%%%%%%%%%%%
When the detection loop antenna is positioned at the center of the $x$-$y$ plane, the obtained transmission spectrum is shown in Fig. 4(b). 
There are two intensity peaks at 12.675 and 13.575 GHz. 
As shown in Figs. 4(c) and 4(d), spatial mapping of the transmission intensity at these frequencies shows clear intensity localization around the center, indicating the hinge modes in Fig. 2(b). 
The strong confinement is consistent with the numerical result shown in Fig. 2(c). 
The frequencies of the two hinge states deviate $\sim$5\% from the numerical results in Fig. 2(b). 
These small deviations are reasonable, considering the fabrication error of the 3D PhC. 

%%%%%%%%%%%%%%%%%%%%%%%%%%
So far, we have introduced the structural shift in the (-$x$, -$y$) quarter part of the second and fourth layers, as shown in Fig. 1(a), and investigated the hinge states in the 3D structure corresponding to the red-dotted square region in Fig. 2(a). 
We next introduce the shift in the (+$x$, -$y$) quarter part of the second and fourth layers, as shown in Fig. 5(a), and fabricate another 3D structure corresponding to the blue-dotted square region in Fig. 2(a) by replacing the second and fourth layers in Fig. 1(a) with the layers in Fig. 5(a). 
We perform MW transmission measurements on this partially replaced 3D PhC following the same measurement procedure as in Fig. 4(a). 
When the detection antenna is placed at the center of the top surface, the resulting transmission spectrum shows no comparable intensity peaks, as shown in Fig. 5(b). 
Also, the 2D transmission intensity map does not show any intensity localization around the center at any frequency. 
These results showing the absence of the hinge modes indicate the selection rule for the hinge modes in this 3D structure. 
From these results, we confirm that the theoretically predicted hinge states in a woodpile structure are experimentally demonstrated in a MW regime. 

%%%%%%%%%%%%%%%%%%%%%%%%%%%%%%%%%%%%%%%%%%%%%%%%%%%%%%%%%%%%%%%%%%%%%%%%%%%%%%%
\section{Summary}
%%%%%%%%%%%%%%%%%%%%%%%%%%%%%%%%%%%%%%%%%%%%%%%%%%%%%%%%%%%%%%%%%%%%%%%%%%%%%%%
We experimentally investigated hinge states in a woodpile PhC based on the theoretical prediction. 
To measure the hinge states in the MW regime, we fabricated a 3D all-dielectric PhC with a millimeter-scale lattice period, consisting of two lattice types that coincide through a structural displacement. 
MW transmission measurements showed spatially localized intensity at one of the two distinct intersections, providing experimental evidence of hinge states. 
In contrast, the transmission spectrum at the other intersection showed no comparable intensity peak, confirming the absence of hinge states as theoretically predicted. 
Our findings of hinge states confined by a large CPBG can be applied to low-loss, metal-free MW transmission lines. 
Furthermore, this work represents an important step towards robust 3D photonic circuits operating in the near-infrared telecommunication bands. 

%%%%%%%%%%%%%%%%%%%%%%%%%%%%%%%%%%%%%%%%%%%%%%%%%%%%%%%%%%%%%%%%%%%%%%%%%%%%%%%
\section{Acknowledgements}\label{sec:acknowledgements}
%%%%%%%%%%%%%%%%%%%%%%%%%%%%%%%%%%%%%%%%%%%%%%%%%%%%%%%%%%%%%%%%%%%%%%%%%%%%%%%
This work was supported by JST CREST Program (Grant Number JPMJCR19T1, Japan) and JST FOREST Program (Grant Number JPMJFR223Q, Japan). H. T. P and K. W. acknowledge the financial support for JSPS KAKENHI (Grants No. JP25K01609, No. JP22H05473, and No. JP21H01019) and Basic Science Research Projects (Grant No. 2401203) from the Sumitomo Foundation.

%%%%%%%%%%%%%%%%%%%%%%%%%%%%%%%%%%%%%%%%%%%%%%%%%%%%%%%%%%%%%%%%%%%%%%%%%%%%%%%

%%%%%%%%%%%%%%%%%%%%%%%%%%%%%%%%%%%%%%%%%%%%%%%%%%%%%%%%%%%%%%%%%%%%%%%%%%%%%%%

\end{document}